\documentclass{aa}  

\usepackage{graphicx}
\usepackage{txfonts}
\usepackage{natbib}
\usepackage{color}
\definecolor{red}{rgb}{1,0,0}
\definecolor{blue}{rgb}{0,0,1}

\begin{document}

 \title{ %NIR flares and sub-mm dips in Sgr~A*:\\Five nights of simultaneous multi-wavelength observations
Flares and variability from Sgr~A* : five nights \\ of simultaneous multi-wavelength observations.
\thanks{Based on multi-wavelength observations obtained at the  ESO VLT Melipal-Yepun telescopes and at the APEX facility (run ID: 179.B-0261).}
        }

 \offprints{X. Haubois}
%   \subtitle{I. Overviewing the $\kappa$-mechanism}

 \author{X. Haubois \inst{1,4}
        \and
         K. Dodds-Eden\inst{2}
        \and
A. Weiss\inst{3}
        \and
        T. Paumard     \inst{1}
	 \and
	       G. Perrin 
        \inst{1}
         \and
        Y. Cl\'enet   
        \inst{1}
         \and
         S. Gillessen \inst{2}
                   \and  
        P. Kervella 
        \inst{1}
               \and  
        F. Eisenhauer\inst{2}
\and
         R. Genzel\inst{2}           
                    \and
D. Rouan\inst{1}
        }

 \institute{LESIA-Observatoire de Paris, CNRS\,UMR\,8109, UPMC Univ Paris 6, Universit\'e Paris Diderot, 5 place Jules Janssen, 92195 Meudon, France 
\\    \email{xavier.haubois@obspm.fr}
 \and MPE, Max-Planck-Institut f\"{u}r Extraterrestrische Physik, Postfach 1312, D-85741 Garching, Germany
\and MPIfR, Max-Planck-Institut f\"{u}r Radioastronomie, Auf dem H\"{u}gel 69, D-53321 Bonn, Germany
 \and   Instituto de Astronomia, Geof\'{i}sica e Ci\^{e}ncias Atmosf\'{e}ricas, Universidade de S\~{a}o Paulo, Rua do Mat\~{a}o 1226, Cidade Universit\'{a}ria, S\~{a}o Paulo, SP 05508-900, Brazil 
%  \email{ste@mpe.mpg.de}        
%  \email{aweiss@mpifr-bonn.mpg.de}
}
 \date{Received 18 July 2011 / Accepted 11 January 2012}

% \abstract{}{}{}{}{} 
% 5 {} token are mandatory

\abstract
 {}
 {We report on simultaneous observations and modeling  of mid-infrared (MIR), near-infrared (NIR) and submillimeter (sub-mm) emission of the source Sgr~A* associated with the supermassive black hole at the center of our Galaxy. Our goal was to monitor the activity of Sgr~A* at different wavelengths in order to constrain the emitting processes and gain insight into the nature of the close environment of Sgr~A*.}
 {We used the MIR instrument VISIR in the BURST imaging mode, the adaptive optics assisted NIR camera NACO, and the sub-mm antenna APEX to monitor Sgr~A* over several nights in July 2007. }
 {The observations reveal remarkable variability in the NIR and sub-mm during the five nights of observation. No source was detected in the MIR but we derived the lowest upper limit for a flare at 8.59$\mu$m (22.4 mJy with $A_{8.59 \mu m}=1.6\pm 0.5$). This observational constraint makes us discard the observed NIR emission as coming from a thermal component emitting at sub-mm frequencies. Moreover, comparison of the sub-mm and NIR variability shows that the highest NIR fluxes (flares) are coincident with the lowest sub-mm levels of our five-night campaign involving three flares. We explain this behavior by a loss of electrons to the system and/or by a decrease in the magnetic field, as might conceivably occur in scenarios involving fast outflows and/or magnetic reconnection.}
 {}%In order to reproduce the multi-wavelength observations, a population of accelerated electrons is needed. n the process, a fraction of the original electrons which "escape"  the system and/or a decrease of the magnetic field. }

 \maketitle
%
%________________________________________________________________

\section{Introduction}

%Accroche
Due to its proximity compared to other galaxies, the center of our Galaxy is the best place to study the nucleus of a spiral galaxy and its strong influence on its neighborhood. In the past two decades, the advent of NIR adaptive optics at 10m class telescopes has made it possible to track the movements of stars in the innermost regions of the Galaxy. The trajectories of these stars, which are perfect Keplerian ellipses with a common focus to measurement precision, have demonstrated there is a supermassive black hole at the center of the Galaxy \citep{1996Natur.383..415E,1997MNRAS.284..576E,1998ApJ...509..678G,2003ApJ...586L.127G,2005ApJ...620..744G,2002Natur.419..694S,2003ApJ...596.1015S}. 

At a distance of about $8.33 \pm 0.35$ kpc, this object has a mass of about 4 $\times$ 10$^{6}$ $M_{\sun}$  \citep{2008ApJ...689.1044G,2009ApJ...692.1075G} located at the position of the compact radio source named Sgr~A* \citep{2004Sci...304..704B}. The Schwarzschild radius of the black hole has an angular size of about 10 $\mu$as as viewed from Earth. This is three times larger than the black holes in M87 \citep[3.66 $\mu$as,][]{2003A&A...399..869B} and in M31 \citep[2.28 $\mu$as,][]{2005ApJ...631..280B}. The GC represents the best evidence so far that there are supermassive black holes in galactic nuclei, making it the best laboratory for testing the supermassive black-hole paradigm and for studying its unique astrophysical environment. 
%Pr?cis?ment les flares  

One of the peculiarities of Sgr~A* is its variability at all wavelengths \cite[see a review in e.g.][]{2010RvMP...82.3121G}. Distinct outbursting events in the lightcurves, called flares, have been detected on different timescales from a few minutes to a few hours, observed in X-rays \citep[e.g.,][]{2001Natur.413...45B,2003ApJ...591..891B} and the NIR \citep[e.g.,][]{2003Natur.425..934G,2004ApJ...601L.159G,2005A&A...439L...9C}. These X-ray/NIR flares are also sometimes followed by a sub-mm flare with a time delay of 100 min to a few hours \citep{2008A&A...492..337E,2009ApJ...706..348Y,2011A&A...528A.140T}. Flares are strongly polarized at the highest flux levels \citep[for a review, see][]{2010RvMP...82.3121G}. Thanks to their detailed polarization analysis of some Sgr~A* flares, \cite{2010A&A...510A...3Z}  support the idea of a compact source for the emission, instead of radially extended shapes.

Moreover, these flares display a quasi-periodicity of about $20$ minutes in the NIR which has been interpreted in terms of the orbiting motion of a hot spot around a rotating (Kerr) supermassive black hole \citep{2003Natur.425..934G,2007MNRAS.375..764T,2006JPhCS..54..443M,2006A&A...455....1E}. Accelerated electrons of Lorentz factor $\gamma \sim 10^3 - 10^6$ are required to produce the NIR and X-ray emission via synchrotron emission and/or inverse Compton scattering \cite[][]{2001A&A...379L..13M,2003ApJ...598..301Y,2004ApJ...606..894Y,2009ApJ...698..676D,2006JPhCS..54..391E}. The possible origins of the hot electrons include magnetic reconnection in the accretion flow \citep{2001Natur.413...45B,2001A&A...379L..13M}, stochastic acceleration \citep{2006ApJ...648.1020L}, fluctuations in the accretion rate, or even the tidal disruption of a small infalling body such as an asteroid \citep{2008A&A...487..527C}. 

While many papers conceive of flares as independent events that are limited in time, other teams claim that the IR emission from Sgr~A* is made up of permanently fluctuating emission that could be described by red noise statistics. The proponents of this picture consider the term ``flare'' as improper  \citep{2009ApJ...691.1021D}. More recently, \cite{DoddsEden2011} has resolved the discrepancy in these two viewpoints, finding that the NIR emission has a dual nature, consisting of two components: the first a low-level, permanently fluctuating component, identified as the NIR ``quiescent'' state, the second consisting of additional, sporadic, high flux ``flare'' events. 

%\rouge{ Marrone 08 : Submillimeter peak is found to occur 100 min after that the NIR flare has been observed }

%Other teams claim that the IR emission can not be called  "flares" and  that it is more likely a statistical fluctuations of the fraction of accelerated electrons that create the observed increase of flux density \citep{2004ApJ...601L.159G,2008arXiv0810.0446D}.

Revealing the enigmatic origin of the flares of Sgr~A* is one of the aims of an ESO large program (LP) of observations (Proposal ID: 179.B-0261, PI : Pr. R. Genzel). This project combined all VLT instruments available for high-resolution infrared studies, as well as sub-mm observations with the Atacama Pathfinder EXperiment (APEX), and was part of a large ground-based and space-based, international multi-wavelength campaign. Constraining the SED of flares by photometry at a high sampling rate in sub-mm, MIR, and NIR can constrain the emission process responsible for the flares. A part of this LP was therefore devoted to observing Sgr A* simultaneously in these three wavelength ranges. 
Throughout this paper we use $\beta$ for the spectral index that is defined through $\nu L_\nu \propto \nu^{\beta}$. For synchrotron emission this can be related to the particle index, $p$, of the synchrotron-emitting electron distribution defined as $n(\gamma) \propto \gamma^{-p}$ through $\beta=(3-p)/2$.

%\rouge{Needs update ?}

In this paper, we analyze simultaneous observations of Sgr A* at NIR, MIR, and sub-mm wavelengths, carried out over five consecutive nights in July 2007. We begin with the observations and data reduction in Sect.~\ref{obs}. In Sect.~\ref{anal}, we present our results. After discussing the implications on the spectral energy distribution (SED) of our observations in Sect.~\ref{model}, we conclude in Sect.~\ref{conc}.

%__________________________________________________________________

\section{Observations and data reduction}
\label{obs}
\begin{figure}
 \centering
 \includegraphics[scale=0.35]{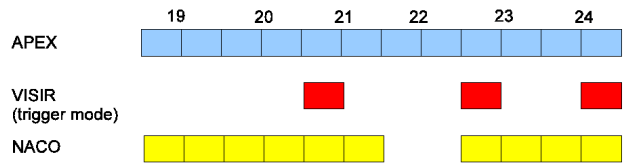}
\caption{Chronogram of the observations}
       \label{log_obs}
 \end{figure}

In this section we present simultaneous multi-wavelength observations carried out during July 2007 at NIR ($2.18\;\mu$m and $3.80\;\mu$m), MIR ($8.59\;\mu$m) and  sub-mm (850~$\mu$m). A schematic log of the observations is presented Fig.~\ref{log_obs}. 

The July 2007 campaign ran for five nights from 18 July 2007 to 23 July 2007. On most nights NACO operated in polarimetric imaging mode in $K$-band, with some short observations at other wavelengths ($H$, $L'$-band), except for July 22 when most of the night was observed in $L'$-band.  A computer crash at APEX on July 18 stopped the sub-mm observations short at $\sim$2:00 UT. On the night of July 21, technical problems with NACO prevented us from obtaining any data, so this night is not included.

\subsection{NIR}
Here we describe the $Ks$ and $L'$-band imaging observations obtained with NAOS-CONICA on UT4 at the VLT \citep{2003SPIE.4841..944L, 2003SPIE.4839..140R}. The data reduction was the same as in \citet{2009ApJ...698..676D} when we subjected the raw data to a sky subtraction computed from jittered object images, followed by flat-fielding and a correction for dead/hot pixels. 

\begin{figure*}
 \centering
 \includegraphics[width=17cm]{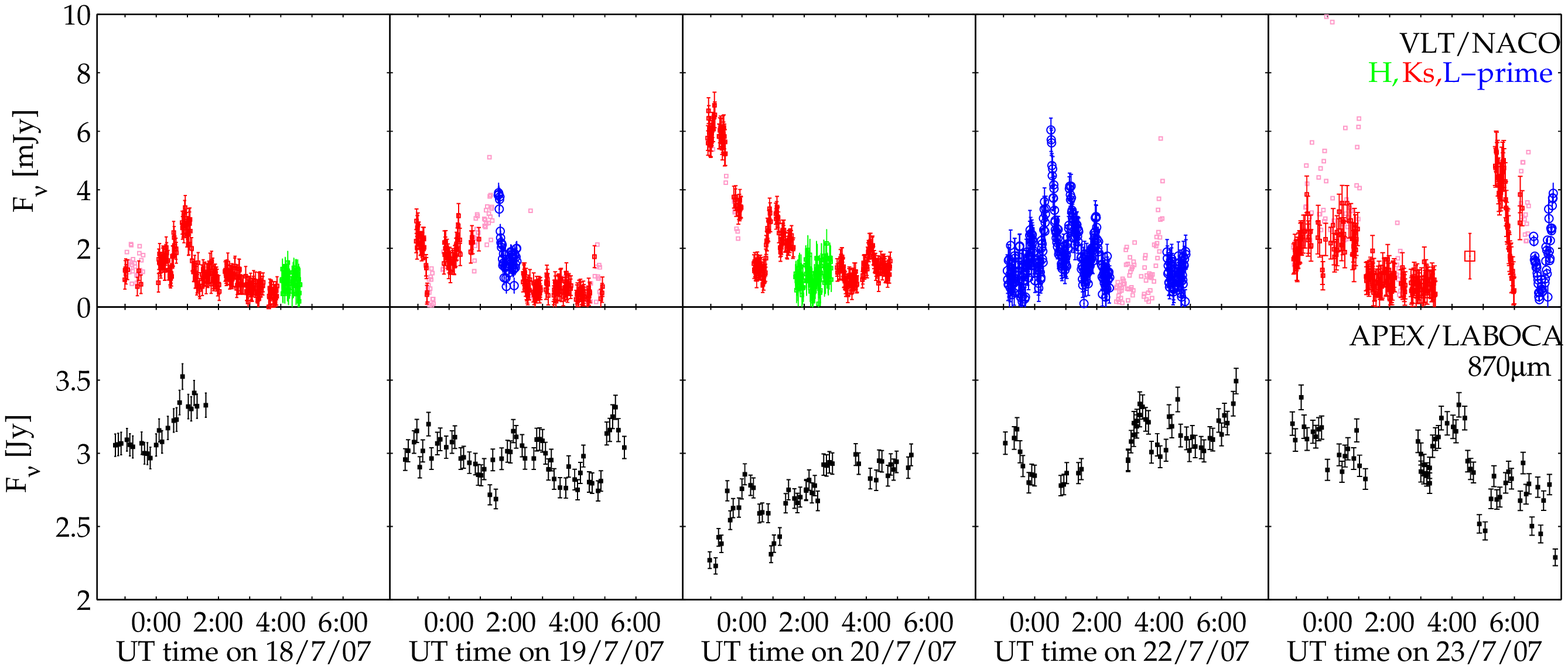}
    \caption{Lightcurves of Sgr~A* at NIR and sub-mm wavelengths for five days of the July 2007 campaign. The green, red, and blue datapoints of the upper panels represent the $H$, $Ks$, and $L'$-band NIR fluxes, respectively. Light red squares show Ks-band observations for which $\max(\mathrm{FWHM}_x,\mathrm{FWHM}_y)>15$\,pixels ($\approx$200 mas). The sub-mm lightcurve is shown below in black. One should bear in mind that the NIR observations were not all made in the same filter.}
       \label{multiplot}
 \end{figure*}

The lightcurves were created with aperture photometry in the same way as in \citet{DoddsEden2011}, using two small apertures, one centered on Sgr A*, the other centered on the close contaminating star S17. At the time of these observations, Sgr A* was confused with the star S17, adding $\approx3.8$ mJy to the $Ks$-band flux \citep{DoddsEden2011}, but only a contribution of $\approx1$ mJy to the $L'$-band flux. We subtracted the contribution due to S17 from the lightcurves shown in Fig.~\ref{multiplot}. However, we note that there may be an additional contribution from faint stars \citep{DoddsEden2011}, and in $L'$-band, perhaps some small contribution from the dust cloud Sgr~A*-f that is very close to Sgr~A*  \citep{2005A&A...439L...9C}. We concentrate only on lightcurves in this paper and leave the polarimetric properties for a future paper. 

The infrared extinction towards the GC has recently been investigated in detail by \cite{2011arXiv1105.2822F}, who derived an accurate extinction law from hydrogen lines, calibrated with a (extinction-free) radio map. We thus used the \cite{2011arXiv1105.2822F}  updated extinctions of $A_{L'}=1.09$ and $A_{Ks}=2.42$ for a red (compared to a stellar source) Sgr A*-like source. 

\subsection{MIR}

MIR observations of Sgr A* at a wavelength of $\lambda=8.59$ $\mu$m were carried out on 20, 22 and 23 July 2007, using the instrument VISIR \citep{2004Msngr.117...12L}, on UT3 at the VLT. These observations were taken in parallel with the NIR observations in a ``trigger'' mode; i.e., whenever increased NIR emission was seen in real time with NACO, VISIR began immediately to observe Sgr A* to increase the likelihood that a flare would be detected in both spectral bands. On July 22 for example, our first VISIR data point on Sgr~A* was obtained around 0:23~UT, about 30~min after the first ($L'$-band) peak, and at the beginning of the brightest, second peak. 

%\subsubsection{Observing in MIR}
We observed in the PAH1 filter centered on 8.59 $\mu$m, the filter with the highest transmission available on VISIR according to the user manual \footnote{http://www.eso.org/sci/facilities/paranal/instruments/visir/doc/index.html}. At these wavelengths, the thermal background emission dominates the images. We used the classical chopping/nodding technique with chopping and nodding throw angles of 10$''$. We chose this parameter because we wanted to continuously monitor the GC on the detector in order to increase the signal-to-noise ratio and to avoid gaps in the temporal coverage. The principle of this chopping/nodding technique is to subtract the high-level sky background (by chopping the secondary mirror) and correct the differential instrumental background caused by the chopping (by nodding the telescope, details can be found in the VISIR user manual).

VISIR was used in BURST mode \citep{2006astro.ph.10322D}. The principle of the BURST mode is to take shorter exposures than the atmospheric coherence time. The main asset of this technique is its spatial resolution. In principle, it allows one to reach the diffraction limit in good observing conditions (0.22$''$ at 8.59 microns, about three times the pixel scale, with an optical seeing $<$0.5$''$),  since the atmospheric turbulence is frozen at the rate at which the images are recorded (an integration time of 20 ms in our case). This gain in angular resolution increases the sensitivity since we have a sharper point spread function (PSF) than with the usual mode. 

\begin{figure*}
  \centering
 \includegraphics[scale=.6]{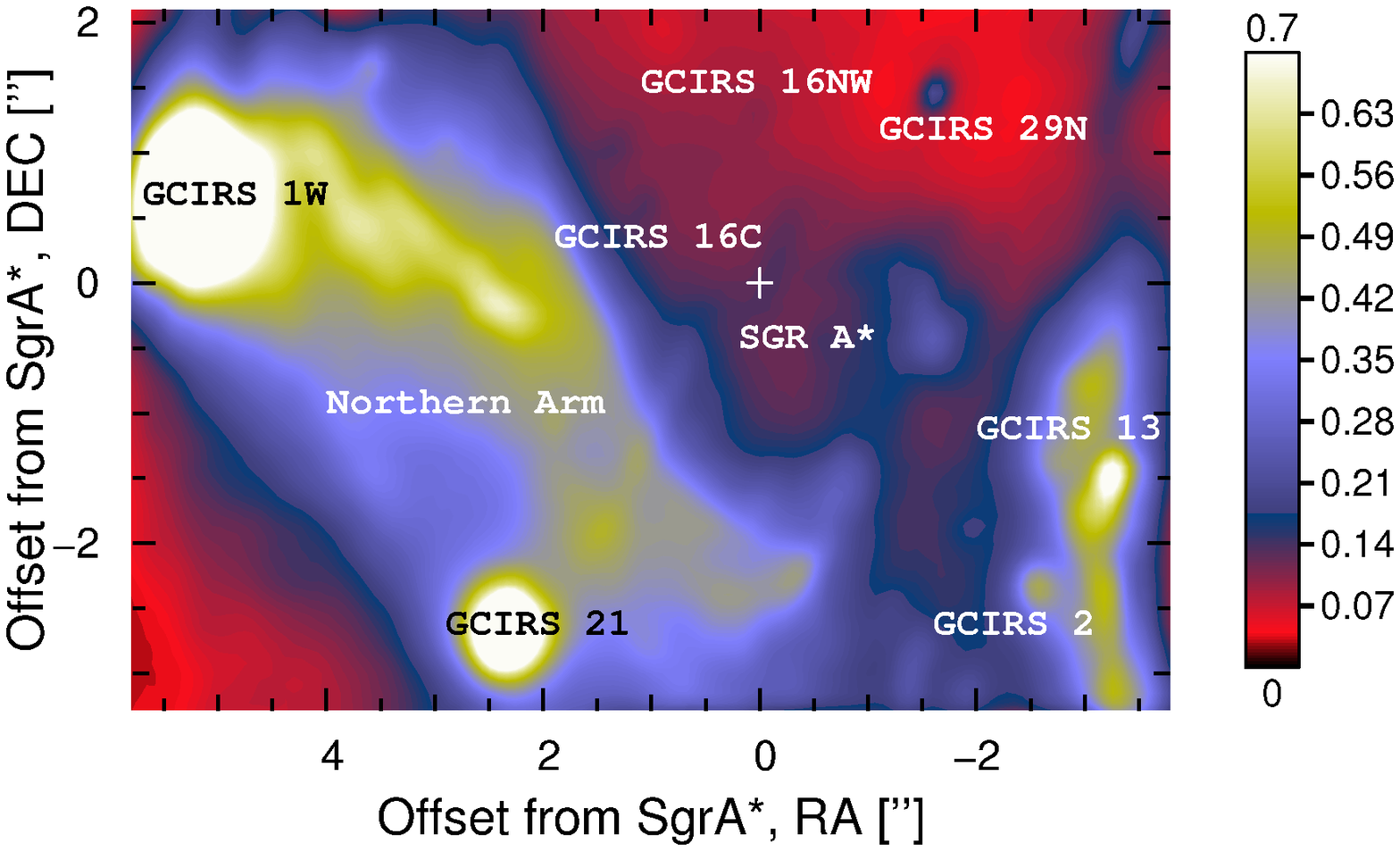}

\vspace{0.7cm}

 \includegraphics[scale=.6]{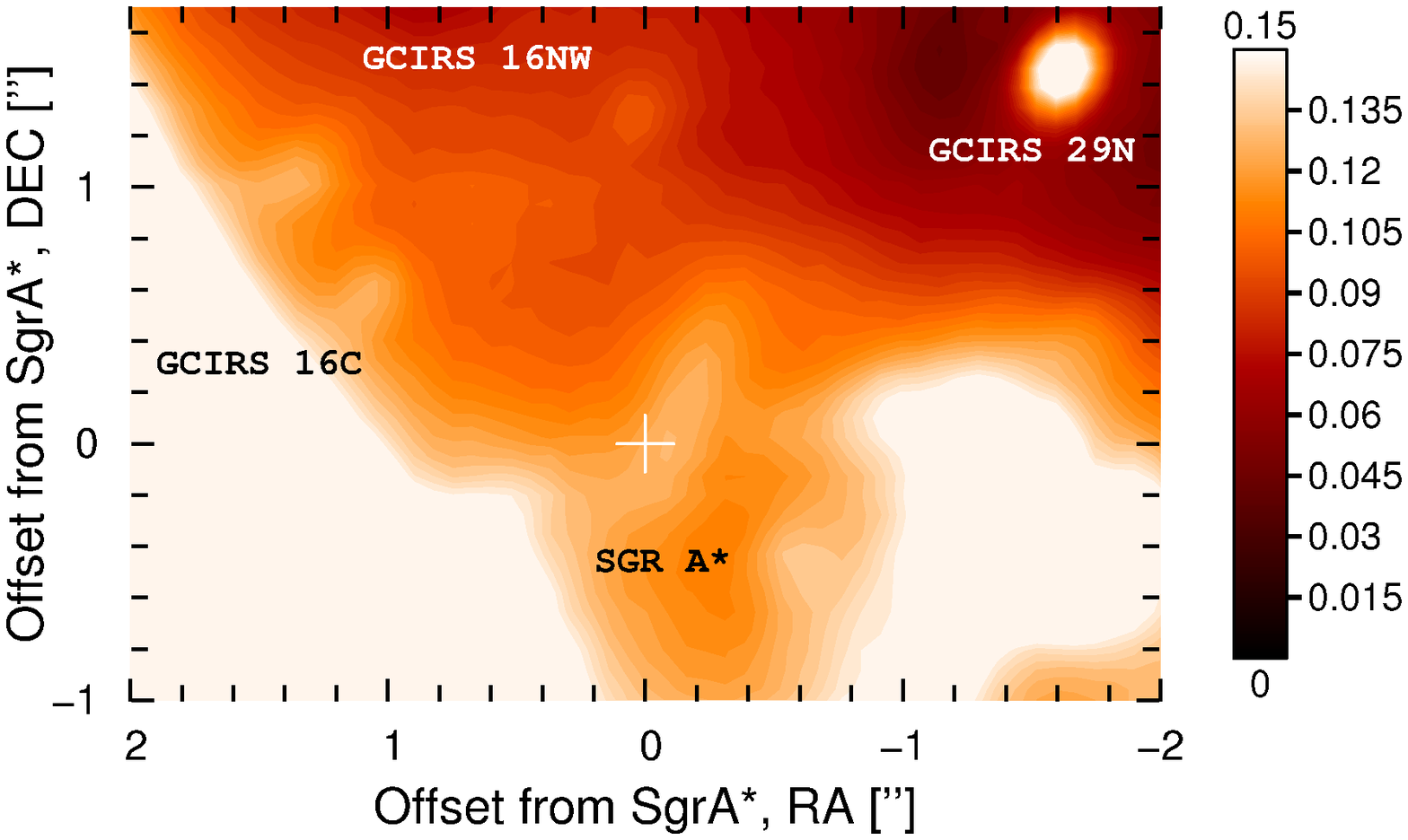}
\caption{Image of the GC zone taken in the PAH1 filter resulting from the co-added images of 3 hours of observation on 22 July 2007. Upper panel: Sgr~A* surroundings compound of stellar sources and gas filaments. Lower panel: close-up on the Sgr~A*  zone. No extinction correction was performed at this point.}
%Retracer les contours...plus clairement, d?convolution ???
         \label{PAH1_image}
  \end{figure*}

%  \subsubsection{Data reduction}
%We reduced the MIR data using a method suitable for the chopping/nodding technique. 
In BURST mode, a typical data set is composed of two nodding cubes (nodding position A and nodding position B). One cube contains 3000 images of the two chopping positions recorded by sequences of 100 frames. The exposure time of one single frame is 20 milliseconds. Thus, a whole cube spans one minute in total. For reduction, the raw frames were first divided by a flat-field calibration image. Using the classical chopping/nodding correction, one obtains four quadrant images with the target appearing as a positive source in two quadrants and a negative source in the other two. However, due to variations in the atmospheric conditions, the image quality varies from frame to frame and a fraction of them are unusable. We thus selected better frames, based on the sharpness of the PSF as estimated on IRS21(which was the only available star to do so in the field of view we observed). Altogether, 85\% of the data are combined in a single cube to increase the signal-to-noise ratio.  Photometric calibration was performed using the observations of several standard stars obtained with our scientific data and the stellar fluxes from \cite{1999AJ....117.1864C}. The final images correspond to about three minutes of accumulated integration time, and they have been co-added in Fig.~\ref{PAH1_image} to show the detected structures better.

%\subsubsection{MIR lightcurve}

We performed aperture photometry at the position of Sgr~A* on all the images taken during the three VISIR half nights to search for variability. The position of Sgr~A* was determined using the positions of the GCIRS stars listed in \cite{2006ApJ...643.1011P} and in \cite{1999ApJ...523..248O}. We applied a low-frequency filter to these lightcurves to detect variability more easily with time scales of about five minutes.  However, the use of the classical chopping/nodding data reduction was unsuitable for constraining the lightcurve of Sgr~A*. Indeed, the lightcurves obtained are different for each quadrant and display a highly noisy pattern because of contamination by field sources (faint stellar or extended diffuse objects) and atmospheric turbulence. As a result we did not perform the classical nodding correction, estimating instead the differential instrument emission (introduced during the chopping phase) from background zones and subtracting it from the Sgr~A* signal. In our case, this technique gives a (sky+instrument) background-corrected flux estimation, free of bias from contaminating sources.

For the 8.59$\mu$m extinction we used $A_{8.59 \mu m}=1.6\pm 0.5$, from  \cite{2011arXiv1105.2822F}  computed for this specific VISIR filter and a red (compared to stars) source spectrum.

\subsection{Sub-mm}

The 870$\mu$m data were taken with the Large
APEX bolometer camera, LABOCA \citep{2009A&A...497..945S}, located on the
Atacama Pathfinder EXperiment (APEX) telescope at Llano Chajnantor,
Chile. LABOCA is an array of 295 composite bolometers operating in the 
345\,GHz transmission window. Its bandwidth is $\sim$ 60 GHz
and the FWHM of the PSF at 870$\mu$m is $~19''$. The atmospheric precipitable water vapor content (PWV) during the 
observations was between 0.6-1.2\,mm corresponding to a zenith opacity
of $0.3-0.45$ for the LABOCA passband.\\

The Galactic Center was observed using the raster-spiral observing
pattern \citep{2009A&A...497..945S}. In this mode the telescope performs a spiral at constant angular speed at four 
positions (the raster positions), leading to a fully sampled map over the full 11$'$ field of view of 
LABOCA. The telescope scanning speed in this mode is between 1$'/s$ and 2.5$'/s$. 
Each spiral takes about 35\,s yielding a total integration time of 140\,s
and a typical rms noise level of $\sim60$\,mJy/beam for each map. The 
Galactic center observations were followed by observations of G10.62, a 
standard LABOCA secondary calibrator, and G5.89 \footnote{See http://www.apex-telescope.org/bolometer/laboca/calibration/}. The atmospheric zenith 
opacity was determined via skydips every one to two hours.\\

The data were reduced using the BoA software package \footnote{BoA: http://www.astro.uni-bonn.de/boawiki/Boa}. Reduction steps on 
the bolometer time series include temperature drift correction based on two 
``blind" bolometers (whose horns were sealed to block the sky signal), 
flat fielding, calibration, opacity correction and correlated noise removal 
on the full array as well as on groups of bolometers related by the wiring 
and in the electronics, flagging of bad bolometers and despiking. Each 
reduced scan was then gridded into a spatial intensity and weighting map. These
reductions step were applied to the Sgr~A* and the G5.89 and 
G10.62 scans. Calibration was achieved using G10.62 which has a flux density
of $33.4\pm2.0$\,Jy for LABOCA \citep{2009A&A...497..945S}. The absolute 
calibration accuracy is about 10\%. To estimate the relative calibration
error we applied the calibration curve of G10.62 to G5.89 and determined
the dispersion of the measured G5.89 fluxes over the observing period
of 7.5\,hours. This yields a relative calibration accuracy of $\sim2$\%.\\
For the determination of the Sgr~A* light curve we first generated
a model of the sub-mm emission in the GC region by co-adding all 
calibrated, pointing-drift-corrected GC maps. From this high signal-to-noise map the point source Sgr A$^*$ was fitted by a Gaussian and 
subtracted from the co-added map. We then reduced all 
GC scans again and subtracted the model signal from the time series of 
each bolometer. The resulting maps only contain the point source Sgr 
A$^*$. The lightcurve was constructed by fitting a Gaussian to each map. To test 
the relative calibration accuracy we also measured the residual peak flux 
at a position near Sgr A$^*$ located on the GC mini-spiral. The residual 
variations were found to be 1.5\% compared to the model flux at this 
position. The LABOCA light curve is shown in Fig.~\ref{multiplot}.

%comparaison SChoedel, Viehmann 2006
%Table des obs :quand ? (comparaison avec NACO) ?

%voir flare note pour la s?lection des images.

\section{Results} \label{anal}

\subsection{High-quality map of GC region at 8.59$\mu$m}

The first interesting outcome of these data are the very high-quality images of the GC area at 8.59 microns.
In Fig.~\ref{PAH1_image}, we see the MIR emission from both the photosphere and the surrounding dust shell of the bright stars GCIRS21 and GCIRS1W.  Also prominent at MIR wavelengths are the gaseous and dusty structures such as the Northern and Eastern Arms of the mini-spiral. To see both Sgr~A* neighborhood and bright stellar and gas components, we present the field of view in two different images.
%and clustered structures in the Eastern Arm \citep{2007A&A...469..993M}
One of the most striking features in these images is  the presence of elongated filaments in the Northern Arm. A proper motion study of these structures in $L'$-band (3.8 $\mu$m)  is reported in \cite{2007A&A...469..993M}. But even more striking is the presence of a stretched emission zone very near Sgr~A* in projection. This structure seems to begin from the Northern Arm (2$''$ southeast of Sgr~A*) and follows an asterism consisting of a few aligned stars leading to the S-star region (those stars are detected in NIR bands). Then, in the most central part (at around 0.5$''$ from Sgr~A*, see lower panel of Fig.÷\ref{PAH1_image}), it deviates a bit to the west, then turns back northwest at the Sgr~A* position (Fig.~\ref{PAH1_image}). This last segment of the structure is also reported in previous papers as a dust ridge whose emission could contaminate a potential point-like emission from Sgr~A* \citep{1996ApJ...470L..45S,2007A&A...462L...1S}.
%While some other filaments can also be matched with  L'-band  observations \citep{2004A&A...424L..21C}, this snake-shaped structure is unveiled for the first time by these MIR data. 
%eckart 2006a dans article schoedel dit que c'est le blob qui est responsable de l'activit? de Sgr?A* en ?tat de qui?tude.
This elongated structure might be due to gaseous material falling towards Sgr~A*  from both sides (southeast and northwest), a bridge of material in projection along the line of sight or a combination of foreground/background components that appear as a unique element in projection. We note that the L'-band source Sgr~A*-f \citep{2005A&A...439L...9C} could belong to this filament, but we cannot conclude anything about the nature of the structure here.  %Even if most of the structures have been identified, some remain ambiguous.  Beyond the fact that we benefited from an unprecedented resolution to image more accurately the already known structures, we possibly detected new sources with a counterpart in the MIR.

\subsection{Variability of Sgr A*}

The NIR and sub-mm lightcurves are shown in Fig.~\ref{multiplot}. In this section we discuss the variability in the NIR (Section \ref{NIRvariability}), an upper limit for detection in the MIR (Section \ref{MIRupperlimit}), and then the sub-mm variability and its relationship with the NIR variability for this campaign (Section \ref{sub-mmvariability}).

\subsubsection{NIR variability}
\label{NIRvariability}

The NIR lightcurves (Fig.~\ref{multiplot}) display significant variability on all five observation nights. If translated to $Ks$-band using a color of $F_\nu \propto \nu^{-0.6}$ \citep{Hornstein2007}, the $L'$-band observations would be fainter by a factor $\approx$0.7; the $H$-band observations would be a factor $\approx1.2$ brighter. For comparison, we computed some statistical quantities to describe the variability of the lightcurves in Table~\ref{stat_NIR}. The strongest flare is seen on July 20 ($Ks$-band; $\approx$7 mJy). On July 22 an interesting variability pattern is seen in the $L'$-band, with four clear peaks separated by $\sim$40 min. This flare was also investigated previously by \cite{2009ApJ...692..902H}, who derived an average period of 45 minutes. This is more than twice as long as the subpeak-to-subpeak timescales observed in previous flares \citep{2003Natur.425..934G,2006JPhCS..54..443M,2006JPhCS..54..391E,2007MNRAS.375..764T,2009ApJ...698..676D}. Moreover, we point out that the flux drops between subpeaks are also very deep. Another flare is seen in the $Ks$-band just after 5:00 UT on July 23.

\begin{table*}
\caption{Statistics of the NIR lightcurves.}             % title of Table
\label{stat_NIR}      % is used to refer this table in the text
\centering                          % used for centering table
\begin{tabular}{c | c c c c c | c c c c c}        % centered columns (4 columns)
\hline\hline                 % inserts double horizontal lines
%\multicolumn{11}{c}{}\\
\multicolumn{1}{c}{}& \multicolumn{5}{c}{NEAR-INFRARED} & \multicolumn{5}{c}{SUBMM} \\
%\multicolumn{1}{c}{}& \multicolumn{5}{c}{} & \multicolumn{5}{c}{} \\ 
\hline
Night (July 2007) & 18 & 19 & 20 &  22 & 23 & 18 & 19 & 20 &  22 & 23 \\  \hline    % inserting body of the table
Median $\pm$ Median deviation (mJy, Jy)  & \multicolumn{5}{c|}{$1.41\pm0.50$} & \multicolumn{5}{c}{$2.96\pm0.14$} \\ %\hline
Median (mJy, Jy) & 0.99 & 1.15 & 1.38 & 1.52 & 1.35 & 3.09 & 2.96 & 2.75 & 3.10 & 2.91 \\ %\hline
Median deviation (mJy, Jy) & 0.27 & 0.61 & 0.46 & 0.56 & 0.67 & 0.09 & 0.11 & 0.15 & 0.11 & 0.18 \\ %\hline
Duration of the sample (min) & 338 & 359 & 351 & 344 & 500 & 174 & 422 & 388 & 444 & 506\\ %\hline
\# peaks above  & 1 & 1 & 2 & 2 & 1 & 0 & 0 & 0 & 0 & 0\\
3-$\sigma$ threshold of entire obs* & & & & &\\  %\hline
\# dips below  & 0 & 0 & 0 & 0 & 0 & 0 & 0 & 2 & 0 & 0\\
3-$\sigma$ threshold of entire obs* & & & & &\\  %\hline
Maximum (mJy,Jy)* avg of the 3 highest points& 3.21  & 3.85  & 6.72 &  5.79 & 5.21 & 3.43 & 3.25 & 2.98 & 3.40 & 3.32 \\ %\hline  
Minimum (mJy,Jy)* avg of the 3 lowest points & -  & -  & - &  - & - & 2.99 &  2.71 & 2.27 & 2.79 & 2.40 \\ %\hline 
Average timespan &  - & - & 120  & 45 & 40 & - & - & 120 & - & -\\
Between peaks/dips (min) & & & &  & \\ \hline
\end{tabular}
\begin{list}{}{}
\item[]Notes: For all nights, the NIR observation were done in the $K$-band except for July 22 where the L' filter was used. We do not consider the $H$-band data here. * A peak/dip is defined as a set of more than 5 points in a row (3 points for sub-mm data) above/below the 3-$\sigma$ level. Flux units are in mJy for NIR, Jy for sub-mm.
\end{list}
\texttt{}\end{table*}

\subsubsection{MIR upper limit}
\label{MIRupperlimit}

The difficulty of detecting Sgr A* in the MIR (compared to the NIR) comes from to the lower spatial resolution and the dominating background.  Despite the trigger mode used to observe with VISIR only when NIR flares were detected, we did not detect any point-like source at the position of Sgr~A* at 8.59 $\mu$m in our observations, nor was there any significant correlation between the NIR and MIR lightcurves obtained via aperture photometry. Up to now, Sgr~A* has never been detected in MIR and only upper limits exist \citep{1995ASPC...73..503T,1996ApJ...470L..45S,2007A&A...462L...1S,2011arXiv1106.5690S}. The lowest upper limit at 8.59$\mu$m to date in the literature gives a mean flux estimation of the Sgr~A* emission of 45+/-13 mJy \citep{2011arXiv1106.5690S}. Once dereddened with the updated 8.59$\mu$m extinction of $A_{8.59 \mu m}=1.6\pm 0.5$ \citep{2011arXiv1105.2822F} computed for the VISIR PAH1 filter, this value becomes 31+/-9 mJy and thus 58 mJy at the 3-sigma level.

We determined an upper limit to the flux by injecting a 3$\times$3 pixels Gaussian-shaped source of known intensity into our MIR images. For each VISIR observation night we subtracted collapsed images taken during intervals of low NIR emission from collapsed image from the flaring state intervals (with the artificially source injected). The detection limit is then derived when this artificial point source is detected at a  3-$\sigma$ significance. 

Because of the high quality of our observations, we are able to derive an especially tight upper limit \emph{on a transient source} at the position of Sgr~A* of 22.4 mJy (dereddened, for July 22), simultaneous to the $L'$-band flare. This is the lowest upper limit ever derived for a flare at 8.59$\mu$m. We emphasize here that, compared to \citep{2011arXiv1106.5690S} who also used this dataset, the upper limit reported here is due to the specificity of our method, which consists in estimating the lowest detectable MIR variability during an NIR flare. This result thus represents an upper limit on the MIR flux relatively to the NIR flux during a flare whereas the Sch{\"o}del et al. one is an absolute MIR upper limit.

Combined with the NIR mean flux in the same time interval, we could also put a lower limit on the spectral index between MIR and the NIR bands. If $\beta$ is defined through $\nu L_\nu \sim \nu^\beta$, the best constraint is obtained for the night of July 23 with a lower limit of $\beta\gtrsim-0.5$  (Table~\ref{upp_limMIR}).

%This represents an unprecedented low detection limit of Sgr~A* emission at 8.59 microns \citep{1996ApJ...470L..45S,2007A&A...462L...1S,1995ASPC...73..503T,2006ApJ...642..861V,2006A&A...455....1E}.

\begin{table*}
\caption{Upper limits on the detection of a synthetic MIR flare, as well as the corresponding limits on the MIR-NIR spectral slope and particle index.}             % title of Table
\label{upp_limMIR}      % is used to refer this table in the text
\centering                          % used for centering table
\begin{tabular}{c c c c}        % centered columns (4 columns)
\hline          % inserts double horizontal lines
%\multicolumn{1}{c}{} &\multicolumn{3}{c}{} \\
%\multicolumn{1}{c}{} &\multicolumn{3}{c}{MIR} \\
%\multicolumn{1}{c}{} &\multicolumn{3}{c}{} \\
\hline 
Night (July 2007) & 20 & 22 & 23  \\   \hline   % inserting body of the table
%Amplitude of the injected gaussian (mJy)  & 5.6 & 5.2  &  7.2 \\
%\hline
On state time interval/s [UT] & 23:07.18 to 00:13:30, & 0:23:12 to 0:39:48,  & 5:34:06 to 5:49:0,\\ 
& \& 00:50:54 to 2:44:54 & \& 1:05:06 to 1:24:48,  & \& 6:04:30 to 6:19:30\\ 
 & & \& 1:57:00 to 2:00:00 & \\% \hline
Off state time interval/s [UT] & 00:17:30 to 00:35.06 & 0:41:06 to 1:03:48, & 5:34:06 to 5:49:06,\\ 
 &  & \& 1:26:12 to 1:55:36,  &  \& 6:04:30 to 6:19:30\\ 
 &  &  \& 2:01:12 to 2:39:12 & \\ %\hline
%Upper limit of a flare detection & 16.3  & 15.1. & 21.0  \\
%at a 3-sigma significance (dereddened)  & & &  \\
%\hline
Upper limit for transient emission & 24.2  & 22.4 & 31.2  \\
(dereddened with $A_{8.59\mu m} =1.6\pm0.5$)  & & &  \\
%\hline
%Average NIR flux (wavelength) & 3.0 (Ks) & 2.3 (L) & 3.6 (Ks) \\
%in on-off intervals& & & \\ \hline
Average NIR flux (wavelength) & 2.0 (Ks) & 3.4 (L$'$) & 4.1 (Ks) \\
on state time intervals& & & \\% \hline
% xav calc: Lower limit on the spectral index &  0.14 & -0.45 &  -0.13  \\
% ($\beta >$ )    & & &  \\  
% old vals: Lower limit on the spectral index &  -0.51 & -1.47 &  -0.56  \\
% ($\beta >$ )    & & &  \\  
Lower limit on the $\nu L_\nu$ spectral index &  $\beta >$-0.80 & $\beta >$-1.3 &  $\beta >$-0.46  \\
($\nu L_\nu \propto \nu^\beta$) & & & \\% \hline
Upper limit on the particle index, p & $p<4.6$ & $p<5.6$ & $p<4.0$ \\
($N(\gamma)\propto \gamma^{-p}$) & & & \\ 
%& & & \\ 
\hline                                   %inserts single line
\end{tabular}
\end{table*}

This constraint is not as strong as (for example) derived by \cite{2009ApJ...698..676D}, who derived a lower limit on the spectral index of $\beta>0.0$ from simultaneous NACO (3.8$\mu$m) and VISIR (11.88$\mu$m) observations. The weaker constraint on the spectral index found here, despite the high sensitivity of the MIR upper limit in this paper, is partly due to the relative faintness of the flares observed during this observation campaign compared to that of \cite{2009ApJ...698..676D} and partly due to the updated extinction law. In fact, the updated extinction also affects the lower limit of \cite{2009ApJ...698..676D}, such that the constraint should be updated to $\beta>-0.3$. Still, both lower limit measurements leave very little room for a negative $\nu L_\nu$ MIR-NIR spectral index.

%Furthermore, our measurements clearly rule out the presence of an exponential cutoff in the NIR for these three flares. Only a thermal component peaking very close to 8.59$\mu$m (which for B=30G would require $\theta_E\gtrsim$2000 with $\theta_E = kT/m\emph{c}^{2}$) could have a spectral index flat enough to satisfy $\beta>-0.5$. It is clear that the observed NIR emission cannot arise from a purely thermal component emitting at \emph{sub-mm} frequencies, or indeed from the high frequency emission from any distribution of electrons with a maximum energy $\gamma_{max} \lesssim$2000, unless the magnetic field is B$>$30G. 
A cutoff at high energy in the flare spectral energy distribution could be produced if the acceleration process is inefficient at accelerating electrons to these energies. If the acceleration process shuts off early, then it can also be due to electron losses from synchrotron cooling. Our measurements clearly rule out that the three flares arise from the high-frequency emission of a spectrum from an electron distribution with an exponential cutoff at sub-mm-emitting energies (i.e. the thermal electrons known to produce the sub-mm emission). While it is hardly disputed that the NIR emission ($\lambda < 4 \mu$m) from Sgr A* requires nonthermal electrons, it is worth illustrating the precise restrictions our data place on the presence of an exponential cutoff in the electron distribution. Assuming such an exponential cutoff in the electron spectrum at an energy of $\gamma_{cutoff}$, with particle index below the cutoff of $p_{0}$,

\begin{equation}
n(\gamma) \sim \gamma^{-p_{0}} \times exp(- \frac{\gamma}{\gamma_{cutoff}})
\end{equation}
the particle index, $p = - \frac{d ln (n(\gamma))}{d ln \gamma}$ will then be $p = p_{0}+\frac{\gamma}{\gamma_{cutoff}}$,
leading to a spectral index $\beta = (3-p)/2 = 1/2 \times (3-p_{0}-\frac{\gamma}{\gamma_{cutoff}}).$

The energy of NIR-emitting electrons is \citep{1979rpa..book.....R}: 
\begin{equation}
\gamma_{NIR} = \sqrt{v_{NIR} \times \frac{16 mc}{3qB}}.
\end{equation}
Thus our measurement of $\beta_{NIR} > -0.5$  for the night of the 23rd implies,
\begin{equation}
\frac{1}{2}(3-p_{0}-\frac{\gamma_{NIR}}{\gamma_{cutoff}}) > - 0.5,
\end{equation}
such that,
\begin{equation}
\gamma_{cutoff} =  \theta_{e}  \gtrsim 200 (\frac{B}{30})^{-1/2} 
\end{equation}
with $\theta_E = kT/m\emph{c}^{2}$, for a thermal distribution (p=2 and $\gamma_{cutoff} =\theta_{e}$), or
\begin{equation}
\gamma_{cutoff} \gtrsim 1200 (\frac{B}{30})^{-1/2}
\end{equation}
for a more ``bottom-heavy" energy distribution below $\gamma_{cutoff}$ (p=3). Thus it is clear that our data require electron energies up to maximum $\gamma$ factors of at least $\gamma_{cutoff} \geqq 200$, and do not allow the NIR flares to arise from a purely thermal distribution of electrons peaking in the sub-mm ($\theta_{e} \sim 10$).

More important, most measurements of the NIR stellar slope between 3.8$\mu$m and 1.65$\mu$m have been independent of photometric calibration and extinction, since the spectral index was measured only relative to stellar colors of the close star S2. Both \cite{Hornstein2007} and \cite{2006ApJ...640L.163G}, for example, found positive slopes of $\approx0.4\pm0.2$ for high fluxes using an `off-state' background subtraction method, which is consistent with the lower limit on the slope reported here, as well as that of \cite{2009ApJ...698..676D}. At fainter fluxes, however, \cite{2005ApJ...628..246E} and \cite{2006ApJ...640L.163G} find a reddening of the spectral index, up to $\beta\approx-3\pm1$ (depending on the background subtraction method) for dim emission of about 2-3 mJy. Our results are, however, inconsistent with results with such steep spectral indices at low fluxes \footnote{Although our observed flares of July 20 and July 22 had peak fluxes comparable with the flare of \cite{2006ApJ...640L.163G}, the interval over which the spectral index was measured had an average NIR flux of only ~ 0.1-0.2 the flux of S2 \citep[and would thus correspond to the ``dim state'' of ][]{2006ApJ...640L.163G}.}. For the three methods of background subtraction presented in \cite{2006ApJ...640L.163G} our results only agree with the ``off-state'' background subtraction method for the dim state, perhaps implying that the other methods underestimate the spectral index. However, we cannot compare the results at low fluxes so easily because our flaring intervals include a mix of both bright and faint NIR emission.

\subsubsection{Submillimeter \& correlation with NIR}
\label{sub-mmvariability}

A striking aspect of the lightcurves presented in Fig.÷\ref{multiplot} is the marked \emph{anticorrelation} seen on the night of July 20, where two peaks in the NIR emission are coincident with the two deepest dips in the sub-mm emission. In particular it is noteworthy that the brightest NIR peak of the five nights of observation is coincident with the lowest sub-mm flux. Similar behavior (an NIR peak accompanying a sub-mm dip) is seen on July 19, as well as on the night of July 23.

\begin{figure}
 \centering
 \includegraphics[width=8cm]{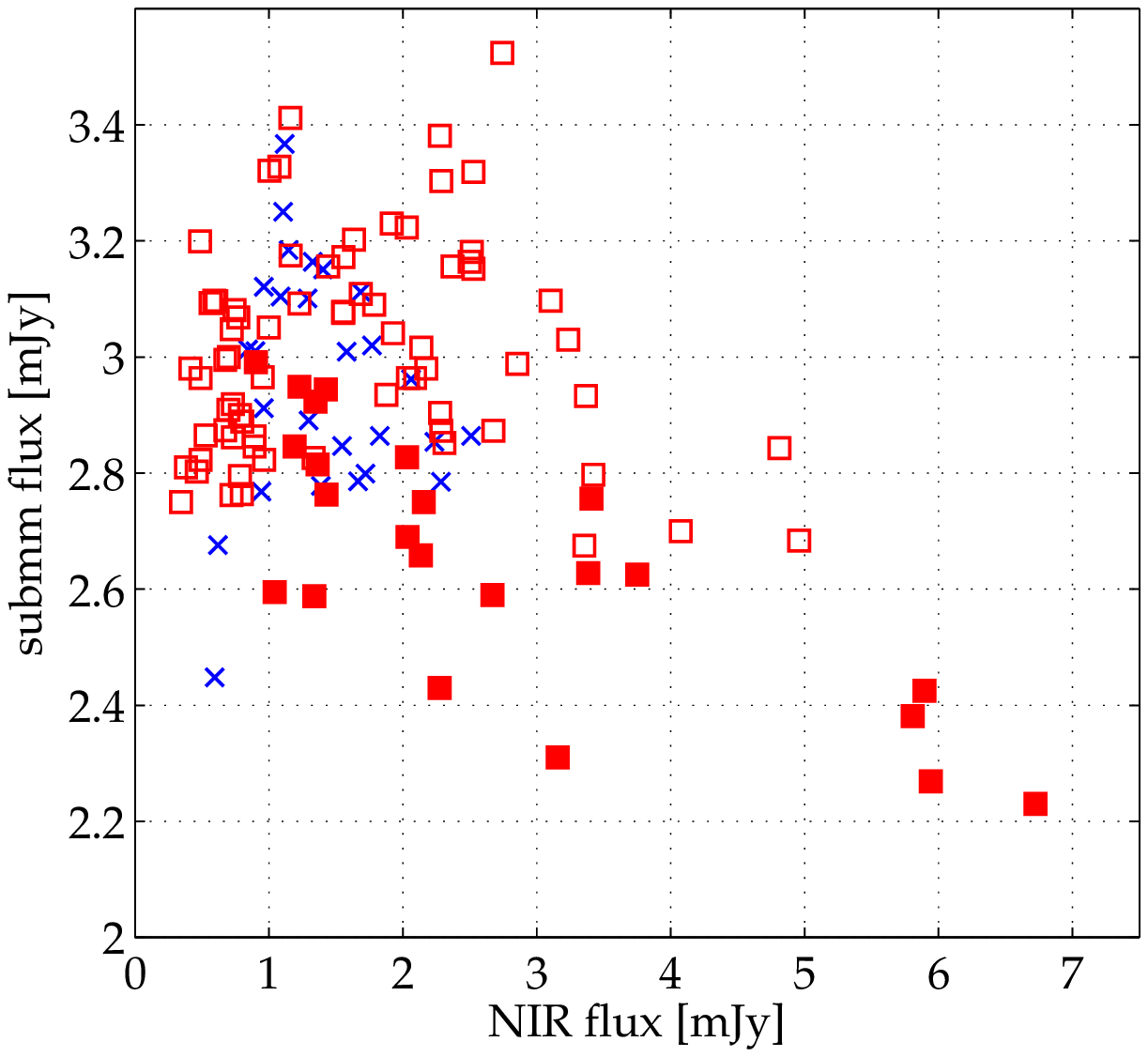}
    \caption{Comparison of simultaneous NIR and sub-mm fluxes for the July 2007 campaign. The graph shows the sub-mm vs. $Ks$-band (red squares) or $L$-band (blue crosses) NIR flux. The July 20 night data are marked as filled-in red squares}. The NIR fluxes were binned to the sampling of the sub-mm lightcurve (5.6 min). The Ks-band scatterplots shows a significant anticorrelation of $r\approx-0.5$, with p-value $p=6.0\times10^{-6}$.
       \label{correlation}
 \end{figure}

There are hints that, a peak in the NIR emission is simultaneous with a peak in the sub-mm emission particularly on July 18. To test whether the apparent anticorrelation of sub-mm and NIR emission in this dataset are significant, we binned the NIR flux to the sampling of the sub-mm lightcurve and compared the bins for non-zero sub-mm and NIR fluxes  (Fig.~\ref{correlation}). Computing the usual (Bravais-Pearson) correlation coefficient of these two variables, an anticorrelation with a coefficient of $r\approx-0.5$ is significant for the $Ks$-band dataset, with a rejection probability of $6.0\times10^{-6}$. The anticorrelation is not significant for the L-band dataset. The shape of the correlation plot shown in Fig.~\ref{correlation} is mostly caused by the anticorrelation of the July 20 flare. If we do not include these data, the correlation is no longer significant and raises the possibility that July 20 lightcurves show an extraordinary behavior, associated with a flare event and not characteristic of the quiescent emission.

A similar relationship between NIR/X-ray and longer wavelength emission was noted by \citet{DoddsEden2010} in observations from April 2007, where a dip in millimeter emission was observed to accompany an NIR/X-ray flare. In that case the brightest NIR/X-ray flux of the campaign (which was indeed a very bright flare, brighter than any in our July campaign) was followed by a $\sim$200 min dip in the millimeter emission. The millimeter flux during the dip was the lowest observed during that entire $\sim$12 day campaign. The explanation put forward in \citet{DoddsEden2010} was that a dip in millimeter emission may occur due to the loss in emissivity of mm-emitting electrons that would occur in a magnetic reconnection event, in which significant magnetic energy is lost in the inner regions of the accretion flow. \cite{2008A&A...492..337E} report a bright NIR flare and a sub-mm signal in which the lowest sub-mm flux is also coincident with the highest NIR flux (between 5h and 6h UT). \cite{2010ApJ...724L...9Y} have recently noted an anticorrelation in NIR/X-ray emission with sub-mm/mm emission, as well in previously published data, extending the number of observed examples to six. The same examples were in most cases originally interpreted as delayed sub-mm flares. This work suggested the anticorrelation could be due to an eclipse of the steady-state accretion flow by the flaring region.

In the following section, we present different modeling attempts to explain this interesting behavior observed in Sgr~A* emission.

\section{SED models for an NIR/X-ray flare with a sub-mm dip}
\label{model}
VLBI measurements show that the size of Sgr A* decreases from radio to the sub-mm \citep[e.g.][]{2004Sci...304..704B,2006ApJ...648L.127B,2008Natur.455...78D}. The sub-mm emission is universally produced in models by hot electrons of $\theta_{e} \sim 10$ in the innermost regions of the accretion flow (close to the last stable orbit), where the electron density is typically $ \sim 10^{7} cm^{-3}$ and magnetic field strength is B$\sim$30G \citep[e.g.][]{2001A&A...379L..13M,2003ApJ...598..301Y}. In the GRMHD simulations of \cite{2010ApJ...717.1092D} and \cite{2009ApJ...706..497M} the mm/sub-mm emission region is situated near the midplane of the disk in the inner few Schwarzschild radii of the accretion flow. Though it is not at all clear where the NIR flares themselves originate, a dip in mm/sub-mm emission simultaneous to an NIR flare implies a causal connection between the two. In what follows, we discuss a number of possibilities for the relationship between the NIR and sub-mm emission region. We consider that the sub-mm dip might occur due to:
%In this section we discuss several possible explanations for the decrease in sub-mm emission observed to accompany the NIR flare on July 20. We divide some possibilities into the following categories, and discuss them in the following sub-sections:
\begin{itemize}
\item a shift in frequency of the sub-mm bump, with the sub-mm emission and NIR emission arising from low and high frequency sides of the sub-mm peak, respectively;
\item a decrease in the total number of sub-mm emitting electrons, through e.g. heating/acceleration (to produce the NIR flare)  accompanied perhaps by cooling/escape of some fraction of the electrons; 
\item a decrease in the sub-mm-emitting electrons emissivity via a decrease in magnetic field \citep{DoddsEden2010};	
\item occultation of the quiescent component by the flaring component \citep{2010ApJ...724L...9Y}. 
\end{itemize}

We do not consider the expanding blob model here, for the simple reason that it does not predict that the highest NIR flux should be coincident with the lowest sub-mm flux.  Furthermore, expanding blob models, which are presented elsewhere in the literature, have the following problems:
% (i) to fit the observations they require a red spectral index in the NIR $\lesssim -0.5$  incompatible with our MIR upper limit, 
(i) some observations appear to require a redder spectral index in the NIR than allowed by our lower limit of $\beta\gtrsim-0.5$ \citep[see e.g. models A and B of Table 4 in][]{2008A&A...492..337E}; the spectral index is given there as $\alpha$ where $S_\nu\propto\nu^{-\alpha}$, so $\alpha = 1 - \beta$,
(ii) Moreover, they require \emph{self-absorbed flare emission in the sub-mm}, which can only occur for the electron densities and/or magnetic fields that are much higher than the typical values in the accretion flow around Sgr A* \citep{DoddsEden2010}.

\subsection{The sub-mm bump}

A natural origin for an anticorrelation between NIR and sub-mm flux is that both high and low frequency parts arise from the same component: the sub-mm bump. %This portion of the spectrum is thought to arise primarily from thermal electrons in the inner regions ($\lesssim10~R_S$) of the accretion inflow/outflow \citep{1997ApJ...490L..77S,1998ApJ...499..731F}. 
One could imagine that if the peak of the sub-mm bump were to vary in frequency, e.g. due to underlying variations in density \citep[through the accretion rate $\dot{M}$ or the magnetic turbulent as shown in][]{2010ApJ...717.1092D}, while essentially preserving the peak sub-mm bump flux, one would observe an increase in the NIR flux as the sub-mm flux decreases (and vice versa).

\begin{figure} 
 \centering
 \includegraphics[width=8cm]{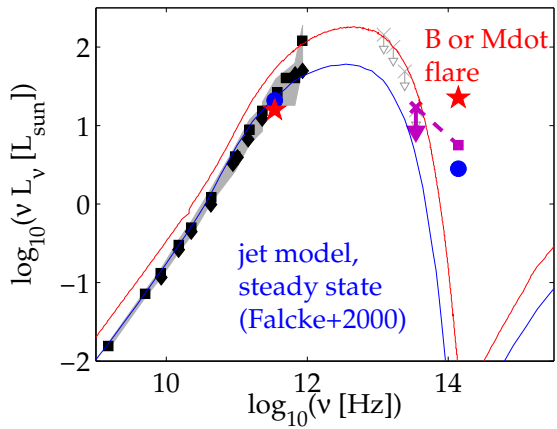}
    \caption{Flare model for Sgr~A* from \cite{2000A&A...362..113F} is plotted in blue and compared to the data from July 20 (purple square for the NIR average flux during the flare time interval and purple arrow for the MIR upper limit; note that the MIR upper limit from the July 22 was lower). The red stars present the peak flux of the July 20 flare, together with the simultaneous sub-mm flux. Blue circles show the median fluxes in a non-flaring state. The radio-sub-mm SED is shown in black (with squares: the compilation of Genzel 2010, diamonds: from Shcherbakov 2011. The typical radio-sub-mm variability amplitudes and a compilation of MIR upper limits are shown in gray and taken from \cite{2001ARA&A..39..309M}. A flare due to a 30\% increase in $B$ in the jet model is also plotted as a red line. An increase in density resulting from a temporary increase in the accretion rate, $\dot{M}$, has the same result \citep{2001A&A...379L..13M}.} 
       \label{fig_jetflares}
 \end{figure}

Such a setup is, however difficult to realize in practice. While an increase in density or magnetic field will increase the frequency at which the flux becomes self-absorbed, it also has the effect of increasing the flux at both sub-mm and NIR frequencies. This is illustrated in Fig.÷\ref{fig_jetflares}, which shows the quiescent relativistic jet model of \cite{2000A&A...362..113F} with a 30\% increase in the magnetic field and an increase in density produced by an increase in $\dot{M}$ \citep{2001A&A...379L..13M}. This model was not created to fit our flare, but is shown more as a demonstration of the general behavior upon increasing $B$ or $n_e$ in the steady-state models. There are two important aspects that are in conflict with our observations and that disfavor this model: (i) the sub-mm flux, as  for the other frequencies, increases with the flare, there is no sub-mm dip; (ii) an exponential cutoff in the flare emission in the NIR is out of the question because of the MIR upper limit: nonthermal particles with at the most $p=4.6$ are required. This reinforces that acceleration of nonthermal electrons is needed to produce an NIR flare \citep{2001A&A...379L..13M,2003ApJ...598..301Y,2009ApJ...698..676D}.

\subsection{Loss of electrons}

Another possibility for a decrease in the sub-mm emission is that a portion of the sub-mm-electrons were accelerated to NIR-emitting energies and/or were lost in an outflow. To determine how many electrons we need for the NIR flare, compared to how many we need to \emph{lose} for the sub-mm dip, we create toy synchrotron models to model both the NIR flare emission as well as for the amount of ``missing'' sub-mm emission. For the details of the models see \cite{DoddsEden2010}. We model the missing emission with a thermal electron population of electron temperature $\theta_e=10$, magnetic field strength $B=30$G, and electron density $n_e$ in a region of size $R$. We then model the flare with the same parameters, but with an accelerated electron distribution, a powerlaw $N(\gamma) \sim \gamma^{-p}$ of $\gamma_{min}=10$ and $p=2.2$ (since the numbers of electrons decrease so rapidly with $\gamma$, the value of $\gamma_{cutoff}$ does not have any effect on the electron density, as long as it is significantly higher than $\gamma_{min}$ and high enough to produce NIR emission). This choice of particle index both matches the published spectral indices for bright emission of $\beta=0.4\pm0.2$ \citep{Hornstein2007,2006ApJ...640L.163G}, satisfies our MIR upper limits, and additionally ensures that the flare itself does not produce too much sub-mm emission (that would result in a simultaneous flare rather than a simultaneous dip).

However, we find that the number of electrons required to produce the ``missing sub-mm emission'' in these models are greater than the number required to produce the flare, if the magnetic field is the same before and after acceleration. Only 30\% of these sub-mm-emitting electrons are required to produce the NIR flare under the conservative assumption $\gamma_{min}=\theta_E$. This implies that the missing sub-mm emission cannot be explained by the acceleration of sub-mm-emitting electrons alone: some fraction must be completely lost to the system in the process.

Given our assumption that electrons are accelerated out of a thermal population, this result is surprisingly robust to the choice of parameters: i.e. $\theta_E$, $B$, for the unaccelerated population of electrons and $\gamma_{min}$, $p$ for the accelerated population. This is because we are essentially only looking at a relative change in the electron distribution. Both the missing sub-mm emission and the flare emission, for example, are subject to the same $B$ by assumption, so changing $B$ essentially only requires a rescaling of the density (which is, however, the same for both unaccelerated and accelerated distributions, so a similar fraction must still escape). The fraction of escaping electrons is more sensitive to $\gamma_{min}$ of the accelerated population. A lower $\gamma_{min}$ however is unrealistic under the assumption of electron acceleration: our condition of $\gamma_{min}=\theta_E$ is already quite conservative since only $\approx$7\% of electrons actually have energies lower than $\theta_E$ in a relativistic Maxwellian distribution (the typical energy is, in fact, $\gamma=2\theta_E$); in other words, electrons with $\gamma<\theta_E$ might as well be treated as escaped for the purposes of this paper. A higher $\gamma_{min}>\theta_E$ only decreases the number of electrons required for the flare (since the density depends sensitively on $\gamma_{min}$), thereby increasing the fraction that must escape. Since it is quite likely that $\gamma_{min}>\theta_E$ \citep{2010ApJ...708.1545D}, we might expect a much more significant fraction of electrons to escape. These electrons could be carried away in a very fast outflow, for example, due to their small number producing very little emission as they rapidly expand and cool \citep[adiabatic expansion][]{DoddsEden2010} or may be accreted onto the supermassive black hole.

\subsection{Decrease in magnetic field}

In \cite{DoddsEden2010}, it was suggested that a decrease in magnetic field \citep[for instance, that accompanies a magnetic reconnection event or due to magnetic turbulence][]{2010ApJ...717.1092D} could cause a decrease in the longer wavelength emission. This could be more efficient at producing a dip in flux than loss of electrons since the synchrotron power emitted by an electron is proportional to $B^2$, while it is only linearly proportional to the number of electrons.

\begin{figure}
 \centering
 \includegraphics[width=8cm]{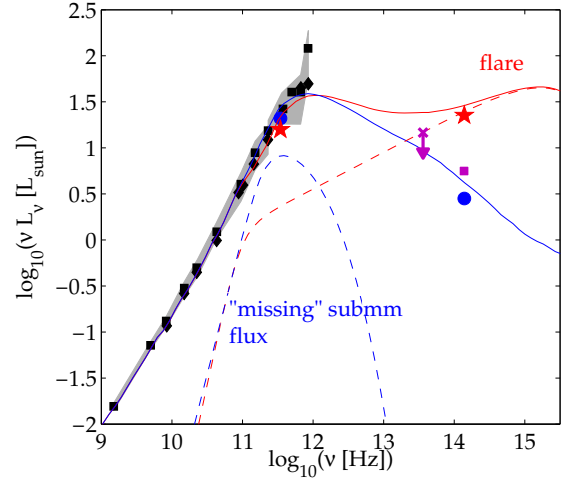}
    \caption{A model for an NIR flare with an accompanying sub-mm dip compared to the data from July 20 (see caption of Fig.÷\ref{fig_jetflares}). Compared to the sub-mm-radio SED shown in black square/diamonds, it can be seen that the flux of the sub-mm dip (red star) is lower than the typical sub-mm flux. The dashed lines show a toy model for the flare, in red, as well as for the ``missing'' sub-mm flux, in blue. The blue solid line shows the quiescent model of \cite{2003ApJ...598..301Y}, and the red solid line represents the same model but with the "missing" sub-mm flux subtracted and the flare emission added. This particular figure shows the decreasing magnetic field case: to prevent overproduction of synchrotron emission from the accelerated population the magnetic field decreases from $B=30$G to $B=14$G. The spectrum for the case of the loss of electrons is very similar.}
       \label{toymodel}
 \end{figure}

In our toy model, we now hold the total number of electrons constant (no escape) between the thermal ($\theta_E\sim10$) and accelerated power-law models ($\gamma_{min}=10$, $p=2.2$) described in the last section. The accelerated power-law model, with the same number of electrons as the thermal model and the same magnetic field strength, overproduces the NIR emission. Instead of decreasing the number of electrons, we find that if we decrease the magnetic field from $B=30$G to $B=14$G, the observations are well-matched. This flare model is shown in Fig.÷\ref{toymodel} compared and combined with the \cite{2003ApJ...598..301Y} model whose thermal component was scaled down to match the SED in the sub-mm part of the section (in particular our blue ``no flare" data point) \footnote{We note that the nonthermal component of the model of \cite{2003ApJ...598..301Y}, with $p=3.5$ lies close to the non-flaring $Ks$-band flux of Sgr A*, which is also similar to the long-term median of the continuously variable NIR flux  \cite{DoddsEden2011}.}.  As in the case of electron loss, $\gamma_{min}=\theta_E$ is quite conservative; if the true $\gamma_{min}$ is higher we would have a higher density of high-energy accelerated electrons, requiring an even greater decrease in the magnetic field to prevent overproduction of NIR emission.  However, the magnetic field in a realistic magnetic reconnection event probably also decreases over a larger region than the acceleration site itself \citep{DoddsEden2010}, which would produce a larger sub-mm dip, while allowing the freedom of a more moderate decrease in $B$.

\subsection{Eclipse}

In \cite{2010ApJ...724L...9Y} it was proposed that a decrease in sub-mm emission could occur via an eclipse of the steady-state emission by the temporary flaring blob. In this section we explore the possibility that occultation is responsible for the sub-mm dips observed in this paper. In this model it is considered that the two regions are separate: i.e. accelerated flare electrons do not come from (or produce very little disturbance to) the quiescent population, and the flaring region is in front of, not embedded in, the quiescent region in the line of sight. 

\begin{figure} 
\centering
\includegraphics[width=8cm]{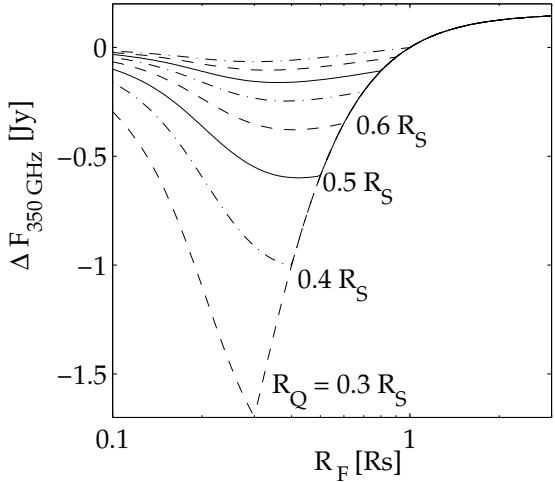}
\caption{Dips in sub-mm emission via occultation by a flaring region of size $R_F$, given a sub-mm flux of $F_{\mathrm{350 GHz}}=3$~Jy, appropriate for our observations of July 20. The different lines show the size of the dip that can be expected for different quiescent region sizes, $R_Q$.}
\label{fig_dipfluxes}
\end{figure}

Figure÷\ref{fig_dipfluxes} shows a plot of the equations of \cite{2010ApJ...724L...9Y} for variation in sub-mm flux vs. size of the flaring region ($R_F$), for different quiescent region sizes ($R_Q$).  Anything larger than $R_Q\gtrsim1 R_S$ encounters very little occultation. Evidently, both the 850$\mu$m and NIR emission regions would have to be very small ($R_Q<0.6R_S$ and $R_F=0.5$-$1.0$ $R_Q$) to produce a dip as large as 0.5 Jy in the sub-mm emission via occultation.

 A quiescent size of $0.5 R_S$ is much smaller than would be expected. Using the caveat from \cite{1983ApJ...264..296M} (radius = 0.9* FWHM) on the FWHM size 3.7$R_S$ (37 microarcseconds) at 1.3mm \citep{2008Natur.455...78D}  and extrapolating with the radius $\propto \lambda^{1.3}$ wavelength dependence  \cite{2006ApJ...648L.127B}, we obtain a radius at 870$\mu$m of 2.0 $R_S$. However, extrapolating the size downwards from 1.3 mm to 850 $\mu$m could be incorrect as the accretion flow might already become optically thin around 1.3 mm. In this case the size would no longer change.

Moreover, the corresponding flare size would be so small that it would imply a very high density of nonthermal particles. It is possible that relaxing the assumption of equipartition in the eclipse model of \cite{2010ApJ...724L...9Y} might change these conclusions. Another possibility is that the electron spectral index might be steeper than the assumed $p=2$ of the model \citep{2010ApJ...724L...9Y}. Although a steeper electron spectrum can increase the optical depth at low frequencies (and thereby the absorption of quiescent emission), it also increases the sub-mm emission from the flare itself significantly \citep[on the order of Jy; ][]{DoddsEden2010}, which is likely to overwhelm any increased absorption of the quiescent sub-mm emission.

% 1 Rs =  2 * G*m/c^2 . Avec m = 4 millions de masses solaires, ca fait 1.18163e+12 cm.

Some radio (43 GHz) dips have been reported as well during an NIR flare \citep{2010ApJ...724L...9Y}. We note that, for the same modeling parameters we used, no significant dips in the radio are expected for any $R_F$, given a 43 GHz ($\lambda$=7mm) emitting region size of $R_Q= 17 R_S$ \citep{2006ApJ...648L.127B}. Only a smaller (43 GHz) quiescent region size can produce observable radio dips due to occultation, for example for $R_Q=4 R_S$ (four times smaller than observed) radio dips of up to $\sim20\%$ of the emission could be produced for flare region sizes of $1-2 R_S$. Yusef-Zadeh et al. argue that a $4\times$ smaller quiescent region size was not necessarily in conflict with the size radio measurements as the source might be much smaller in semi-minor axis; however, this in turn imposes increasingly difficult constraints on both the geometrical shape of the flaring region and its alignment with respect to the quiescent region to retain a high degree of occultation (given that a covering fraction $(R_F/R_Q)^{2}$  of at least $0.5^2$ is required).

\section{Conclusion}
\label{conc}

We have reported on a sub-mm, NIR, and MIR observation campaign of Sgr~A* over five nights in July 2007. First, we detected NIR flare events of different amplitudes and structures. We also estimated the lowest upper limit ever derived for a flare at 8.59$\mu$m (22.4 mJy with $A_{8.59 \mu m}=1.6\pm 0.5$) and derived constraints on the spectral slopes between NIR and MIR. The analysis of the correlation between sub-mm and NIR variability shows that the highest NIR brightness levels are coincident with the lowest sub-mm levels. We examined several possible explanations for this observational fact. We disfavor variations in the sub-mm bump mainly because the sub-mm and NIR emission from only one thermal component does not satisfy our MIR upper limit. Occultation of the flaring component by the quiescent component \citep{2010ApJ...724L...9Y} seems unlikely given the small sizes of the emitting areas at play. We argue that the NIR/sub-mm anticorrelation could  likely be due to either electron loss ($>$70\% for $\gamma_{min}=\theta_E$ ), a decrease in the magnetic field (from $B=30$G to $B=14$G), or a combination of both, involving fewer escaping particles and a smaller decrease in magnetic field. It is worth noting that a decrease in magnetic field, as expected during a magnetic reconnection event, can also explain the different durations of observed flares in the NIR and X-ray lightcurves \citep{DoddsEden2010}. Regardless of the mechanism, we conclude that there is a link between a decrease in sub-mm emission and the acceleration of nonthermal electrons (or thermal electrons with $\theta_E >$2000). While more work is needed to pin down the nature of this link precisely, magnetic activity, fast outflows, and their interaction with the black hole are very likely to be pieces of the puzzle.

\begin{acknowledgements}
This work was supported by a grant from R{\'e}gion Ile-de-France. We also received the support of PHASE, the high angular resolution partnership between the ONERA, Observatoire de Paris, CNRS, and Paris Diderot University.

\end{acknowledgements}

\end{document}